\documentclass[11pt,twoside,onecolumn]{article}
\usepackage[]{latexsym}
\usepackage{epsfig}
\usepackage{amsmath,amssymb}
\RequirePackage[dvipsnames,usenames]{color}
\newcommand{\be}{\begin{eqnarray}}
\newcommand{\ee}{\end{eqnarray}}

\title{\bf Searching for modified gravity: a conformal sector?}
\author{J Ovalle$^{ab}$\thanks{jovalle@usb.ve; Based on the contributed lecture given at the  IF-YITP Cosmo International Symposium. Phitsanulok, Thailand, August 3-5, 2016.}
$\,$ 
R Casadio$^{cd}$\thanks{casadio@bo.infn.it}
$\,$ 
A Sotomayor$^{e}$\thanks{adrian.sotomayor@uantof.cl}
\\
\null
\\
$^a${\em The Institute for Fundamental Study, Naresuan University}
\\
{\em Phitsanulok 65000, Thailand}
\\
$^b${\em Departamento de F\'{\i}sica, Universidad Sim\'on Bol\'ivar,}
\\
{\em AP 89000, Caracas 1080A, Venezuela}
\\
$^c${\em Dipartimento di Fisica e Astronomia, Alma Mater Universit\`a di Bologna}
\\
{\em via Irnerio~46, 40126 Bologna, Italy}
\\
$^d${\em Istituto Nazionale di Fisica Nucleare, Sezione di Bologna, I.S.~FLAG}
\\
{\em  viale Berti~Pichat~6/2, 40127 Bologna, Italy}
\\
$^e${\em Departamento de Matem\'aticas, Universidad de Antofagasta}
\\
{\em  Antofagasta, Chile}
}
\begin{document}
\maketitle
\begin{abstract}

We conjecture that any modification of general relativity can be studied by the minimal geometric deformation approach provided that such modification can be represented by a traceless energy-momentum tensor.
\end{abstract}
%
%
%
%
%
%
%
\section{Introduction}
As is well known, there are some fundamental questions associated with the gravitational interaction
which general relativity (GR) cannot answer satisfactorily, namely the dark matter problem and the introduction of dark energy to explain the accelerated expansion of the universe. Moreover, the incompatibility between GR and the Standard Model of particle physics, 
or equivalently, the failure to quantize GR by the same successful scheme used with the other
fundamental interactions.
Such issues have motivated the search for new gravitational theories beyond GR that could help
to explain part of these problems. Among these alternatives, we have extra-dimensional theories. In this context the minimal geometric deformation approach (MGD), originally proposed~\cite{jo1} in the the Randall-Sundrum brane-world~\cite{lisa1,lisa2}, was developed and later extended to investigate new black hole
solutions~\cite{MGDextended1,MGDextended2} (For a recent brief review, see~\cite{MGDbrief}).
While the brane-world is still an attractive scenario, to find interior solutions for self-gravitating systems is a difficult task,
mainly due to the existence of non-linear terms in the matter fields. Despite these complications, the MGD has proven to be useful, among other things, to derive
exact and physically acceptable solutions for spherically symmetric and non-uniform stellar
distributions~\cite{jo2,jo5} as well; to express the tidal charge in the metric found in
Ref.~\cite{dmpr} in terms of the usual Arnowitt-Deser-Misner (ADM) mass~\cite{jo6};
to study microscopic black holes~\cite{jo7}; to clarify the role of exterior Weyl stresses acting on
compact stellar distributions~\cite{jo8,jo9}; to extend the concept of variable tension
introduced in Refs.~\cite{gergely2009} by analyzing the shape of the black string in the
extra dimension~\cite{jo10}; to prove, contrary to previous claims, the consistency
of a Schwarzschild exterior~\cite{jo11} for a spherically symmetric self-gravitating
system made of regular matter in the brane-world; to derive bounds on extra-dimensional
parameters~\cite{jo12} from the observational results of the classical tests of GR in the
Solar system;
to investigate the gravitational lensing phenomena beyond GR~\cite{roldaoGL}, to determine the critical stability region for Bose-Einstein condensates
in gravitational systems~\cite{rrplb}, as well as to study the corrections to dark SU(N) stars observable parameters due to variable tension fluid branes~\cite{SUN}.
\section{The MGD: a qualitative approach}
\label{s2}
Let us start by assuming that there is a ``superior" gravitational theory, superior in the sense that it could explain the dark matter problem, what dark energy is, and maybe to explain the quantum problem. Of course, this new gravitational theory must contain GR at some sector, that is, it should be capable of reproducing GR at solar system scale for instance. The reason for this is the fact that GR is a well tested and successful theory, at least at solar system scale. This new gravitational theory can be seen as GR plus some ``corrections" These corrections always produce consequences on the energy-momentum tensor $T_{\mu\nu}$. Indeed, these corrections can be consolidated as part of an effective energy-momentum tensor $\theta_{\mu\nu}$ in such a way that 
\begin{equation}
T_{\mu\nu}\rightarrow\,T_{\mu\nu}+\alpha\theta_{\mu\nu}\ ,
\end{equation}
where $\alpha$ is a free parameter associated with the new gravitational sector not described by GR. Of course, when we turn off this free parameter we return to the domain of GR. This limit represents not only a critical point for a consistent extension of GR, but also a non-trivial problem
that must be treated carefully. In this respect, one cannot try and change GR without considering the well-established and very useful
Lovelock's theorem~\cite{lovelock}, which severely restricts any possible ways of modifying
GR in four dimensions. We will now show the simplest generic way.
\par
The simplest way to extend GR is by considering a modified Einstein-Hilbert action,
\begin{equation}
\label{corr1}
S
=
\int\left[\frac{R}{2\,k^2}+{\cal L}\right]
\sqrt{-g}\,d^4x
+\alpha\,({\rm correction})
\ ,
\end{equation} 
where the generic correction shown in Eq.~(\ref{corr1}) should be, of course,
a well justified and physically motivated expression.
At this stage the GR limit, obtained by setting $\alpha = 0$, is just a trivial issue, so everything looks
consistent.
Indeed, we may go further and calculate the equations of motion by the usual way, hence
\begin{equation}
\label{corr2}
R_{\mu\nu}-\frac{1}{2}\,R\, g_{\mu\nu}
=
k^2\,(T_{\mu\nu}+\alpha\theta_{\mu\nu})
\ .
\end{equation}
The new term  $\theta_{\mu\nu}$ in Eq.~(\ref{corr2}) in general has new fields, like scalar, vector and tensor fields, all of them coming from the new gravitational sector not described by Einstein's theory.
At this stage the GR limit, again, is a trivial issue, since $\alpha = 0$ leads to the standard Einstein's equations 
$G_{\mu\nu}=k^2\,T_{\mu\nu}$. However, when the system of equations given by the expression~(\ref{corr2}) is solved, in general the solution eventually found cannot reproduce the GR limit by
simply setting $\alpha = 0$.
The cause of this problem is the non-linearity of Eq.~(\ref{corr2}), and should not be a surprise.
To clarify this point, let us consider a generic spherically symmetric solution for the radial metric component, namely 
\begin{equation}
\label{g11-1}
g^{-1}_{rr} = 1 - \frac{2\, m(r)}{r}
\ ,
\end{equation}
where $m$ is the mass function of the self-gravitating system. Now let us consider the effect of the new gravitational sector in~(\ref{corr1}), which modifies the equation of motion in Eq.~(\ref{corr2}) by the new term  $\theta_{\mu\nu}$. This new term produces a ``geometric deformation" on the radial metric component~(\ref{g11-1}), which generically may be written as
\begin{equation}
\label{g11def}
g^{-1}_{rr} = 1 - \frac{2\, m(r)}{r} + ({\rm geometric\ deformation})
\ ,
\end{equation}
where by {\it geometric deformation\/} one should understand  a deviation from the generic GR solution in~(\ref{g11-1}). It is now very important to note that the deformation~(\ref{g11def}) always takes the form
\begin{equation}
\label{def}
({\rm geometric\ deformation}) = X + \alpha\,Y
\ .
\end{equation}
This expression is very significant, since it shows that the GR limit cannot be  {\em a posteriori\/}
recovered by setting $\alpha=0$, since there is a ``sector'' denoted by $X$ which does not depend on $\alpha$.
Consequently, the GR solution is not trivially contained in this extension,
and one might say that we have an extension to GR which does not contain GR.
A method that solves the non-trivial issue of consistency with GR described above is the so-called
{\it Minimal Geometric Deformation} MGD approach~\cite{jo1}. 
The idea is to keep under control the anisotropic consequences on GR appearing in the extended theory,
in such a way that the $\alpha$-independent sector in the geometric deformation shown as $X$
in Eq.~(\ref{def}) always vanishes.
This is accomplished when a GR solution is forced to remain a solution in the extended theory.
Roughly speaking, we need to introduce the GR solution into the new theory, as far as possible.
This provides the foundation for the MGD approach.
\section{MGD and $f(R)$ theories}
\label{s3}

We know the MGD works very well to generate new brane-world solutions from already known GR solutions. Since the MGD does not care about the source of the modification, in the sense that these modifications always can be consolidated as an energy-momentum tensor $\theta_{\mu\nu}$, there is no reason to believe it will not work properly in any other extension beyond the brane-world. In this respect, it would be interesting to test this approach in the context of $f(R)$ theories. The reason is that these theories have interesting properties~\cite{diFelice,thomas,salvatore,salvatore2,marcelo1,marcelo2} and appears in a natural way in superstring theories. 
\par
When we use the MGD in $f(R)$ theories, we found that the geometric deformation $g^{*}(r)$ undergone by the Schwarzschild metric, namely
\begin{equation}
\label{Schw}
ds^2=\left(1-\frac{2{\cal M}}{r}\right)dt^2-\left(1-\frac{2{\cal M}}{r}+\alpha\,g^{*}(r)\right)^{-1}dr^2-d\Omega^2\ .
\end{equation}
yields to the trivial solution $g^{*}(r)=0$. This is unexpected since we know there are solutions beyond the Schwarzschild metric in the spherically symmetric vacuum under $f(R)$, hence we should at least to reproduce some of these solutions. We can conclude thus that the MGD in its original form and $f(R)$ are incompatible. Therefore we need to modify the MGD approach to make it compatible with $f(R)$, but there are not many options to accomplish this. Indeed, the only possible extension is to introduce by hand a deformation on the temporal metric component, as developed in Ref.~\cite{MGDextended1}. When this extension is used in $f(R)$ we end up with a very complicated system of equations. However, the model $f(R)=R + \alpha\,R^2$ produces a simpler system which could be useful to find new vacuum solution beyond the Schwarzschild metric. This should not be a surprise, since it is well known that the pure $f(R)=R^2$ is conformally equivalent to Einstein gravity with a minimally coupled scalar field and a cosmological constant \cite{Alvarez}, so we should anticipate that in this particular case the MGD and $f(R)$ seems to be compatible with the searching for new vacuum solutions. At this stage we can ask ourselves what do the effective Einstein field equations in the brane-world and $f(R)=R^2$ have in common? In the case of the brane-world, if high-energy corrections are removed keeping only non-local corrections, then the energy-momentum tensor representing the new gravitational sector is traceless. Hence it can be associated with a conformal gravitational sector. We know that in this case, namely, the brane-world with only non-local corrections, the MGD works very well. All this tell us that any modification of general relativity can be studied by the MGD provided that such modification can be represented by a traceless energy-momentum tensor, or equivalently, associated with a conformal gravitational sector.

\begin{figure}[t]
\center
\includegraphics[scale=0.3]{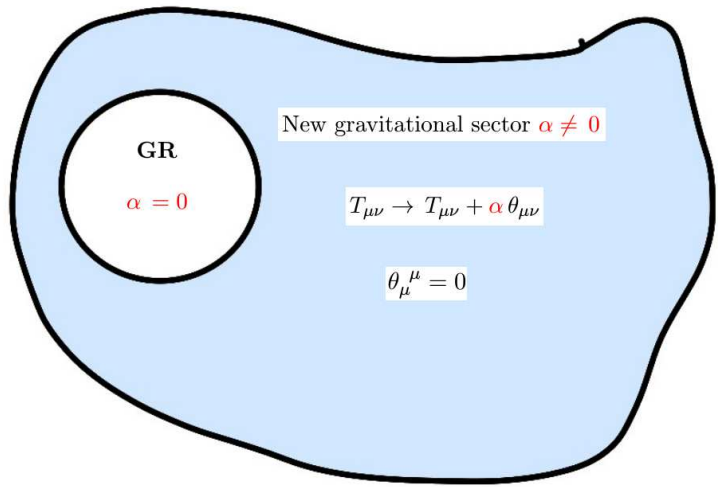}
\\
\centering\caption{When the energy-momentum tensor $\theta_{\mu\nu}$ associated with the 
new gravitational sector is traceless, the MGD works properly. It seems to show a link between conformal invariance and the MGD.}
\label{pre}      
\end{figure}
\section{Acknowledgements}
A.S. is partially supported by Project Fondecyt 1161192, Chile.

\end{document}